\documentclass[aps,prd,a4paper,amsmath,amssymb,nofootinbib,showpacs,showkeys,
preprintnumbers,notitlepage,superscriptaddress,twocolumn]{revtex4-1}

\usepackage{bm}
\usepackage[dvips]{color,graphicx}

\begin{document}

\title{Deconfinement and Chiral Symmetry Restoration in a Strong Magnetic Background}

\author{Raoul Gatto}\email{raoul.gatto@unige.ch}
\affiliation{Departement de Physique Theorique,
 Universite de Geneve, CH-1211 Geneve 4, Switzerland}

\author{Marco Ruggieri}\email{ruggieri@yukawa.kyoto-u.ac.jp}
\affiliation{Yukawa Institute for Theoretical Physics,
 Kyoto University, Kyoto 606-8502, Japan}



\begin{abstract}
We perform a model study of deconfinement and chiral symmetry
restoration in a strong magnetic background. We use a Nambu-Jona
Lasinio model with the Polyakov loop, taking into account a
possible dependence of the coupling on the Polyakov loop
expectation value, as suggested by the recent literature. Our main
result is that, within this model, the deconfinement and chiral
crossovers of QCD in strong magnetic field are entangled even at
the largest value of $eB$ considered here, namely $eB=30 m_\pi^2$
(that is, $B \approx 6\times 10^{15}$ Tesla). The amount of split
that we measure is, at this value of $eB$, of the order of $2\%$.
We also study briefly the role of the 8-quark term on the
entanglement of the two crossovers. We then compare the phase
diagram of this model with previous results, as well as with
available Lattice data.
\end{abstract}

\pacs{12.38.Aw,12.38.Mh}\keywords{Hot quark matter, Effective
models of QCD, Deconfinement and chiral symmetry restoration in
magnetic fields.} \preprint{YITP-10-96}\maketitle

\section{Introduction}
The study of the Quantum Chromodynamics (QCD) vacuum, and of its
modifications under the influence of external factor like
temperature, baryon chemical potential, external fields, is one of
the most attractive topics of modern physics. One of the best
strategies to overcome the difficulty to study chiral symmetry
breaking and deconfinement, which share a non-perturbative origin,
is offered by Lattice QCD simulations at zero chemical potential
~\cite{deForcrand:2006pv,Aoki:2009sc,Bazavov:2009zn,Cheng:2009be,Karsch:2000kv}.
At vanishing quark chemical potential, it is established that two
crossovers take place in a narrow range of temperature; one for
quark deconfinement, and another one for the (approximate)
restoration of chiral symmetry. Besides, the use of the
Schwinger-Dyson equations for the quark
self-energy~\cite{Fischer:2009wc,Aguilar:2010cn}, and the use of
functional renormalization group~\cite{Wetterich:1992yh} to the
Hamiltonian formulation of Yang-Mills theory in Coulomb gauge, are
very promising~\cite{Feuchter:2004mk,Leder:2010ji}. QCD with one
quark flavor at finite temperature and quark chemical potential
has been considered in~\cite{Braun:2008pi} within the functional
renormalization group approach, combined with re-bosonization
tecniques; in~\cite{Braun:2009gm}, the renormalization group flow
of the Polyakov-Loop potential and the flow of the chiral order
parameter have been computed. The results of~\cite{Braun:2009gm}
suggest that the chiral and deconfinement phase transition
temperature agree within a few MeV for vanishing and (small) quark
chemical potentials. Furthermore, an interesting possibility to
map solutions of the Yang-Mills equations of motion with those of
a scalar field theory has been proved in~\cite{Frasca:2007uz}.

An alternative approach to the physics of strong interactions,
which is capable to capture some of the non-perturbative
properties of the QCD vacuum, is the Nambu-Jona Lasinio (NJL)
model~\cite{Nambu:1961tp}, see Refs.~\cite{revNJL} for reviews. In
this model, the QCD gluon-mediated interactions are replaced by
effective interactions among quarks, which are built in order to
respect the global symmetries of QCD. Beside this, the parameters
of the model are fixed to reproduce some phenomenological quantity
of the QCD vacuum; therefore, it is reasonable that the main
characteristics of its phase diagram represent, at least
qualitatively, those of QCD.

In recent years, the NJL model has been improved in order to be
capable to compute quantities which are related to the
confinement-deconfinement transition of QCD. It is well known that
color confinement can be described in terms of the center symmetry
of the color gauge group and of the Polyakov
loop~\cite{Polyakovetal}, which is an order parameter for the
center symmetry in the pure gauge theory. In theories with
dynamical fermions, the Polyakov loop is still a good indicator
for the confinement-deconfinement transition, as suggested by
Lattice data at zero chemical
potential~\cite{deForcrand:2006pv,Aoki:2009sc,Bazavov:2009zn,Cheng:2009be,Karsch:2000kv}.
Motivated by this property, the Polyakov extended Nambu-Jona
Lasinio model (P-NJL model) has been
introduced~\cite{Meisinger:1995ih,Fukushima:2003fw}, in which the
concept of statistical confinement replaces that of the true
confinement of QCD, and an effective potential describing
interaction among the chiral condensate and the Polyakov loop is
achieved by the coupling of quarks to a background temporal gluon
field, and then integrating over quark fields in the partition
function. The P-NJL model, as well as its renormalizable
extension, namely the Polyakov extended Quark-Meson model (P-QM),
have been studied extensively in many
contexts~\cite{Ratti:2005jh,Roessner:2006xn,Megias:2006bn,Sasaki:2006ww,
Ghosh:2007wy,Fukushima:2008wg,Abuki:2008nm,Sakai:2008py,Sakai:2009dv,
Abuki:2008tx,Hell:2008cc,Kashiwa:2007hw,Sakai:2010rp,Herbst:2010rf,
Kahara:2008yg,Partyka:2010em,Bhattacharyya:2010wp,Skokov:2010uh}.

In a remarkable paper~\cite{Kondo:2010ts} it has been shown by
Kondo that it is possible to derive the effective 4-quark
interaction of the NJL model, starting from the QCD lagrangian. In
his derivation, Kondo has shown explicitly that the NJL vertex has
a non-local structure, that is, it is momentum-dependent; besides,
the vertex acquires a non-trivial dependence on the phase of the
Polyakov loop. This idea has been implemented within the P-NJL
model in~\cite{Sakai:2010rp}; the modified model has then been
named EPNJL, and the Polyakov-loop-dependent vertex has been
called entanglement vertex. We will make use of this nomenclature
in the present article. Before going ahead, it worths to notice
that a low-energy limit of QCD, leading to the non-local
Nambu-Jona Lasinio model studied in~\cite{Langfeld:1996rn}, has
been discussed independently in~\cite{Frasca:2008zp}. In this
reference, the low-energy limit of the gluon propagator leads to a
relation among the NJL coupling constant and the string tension.

In this article, we report on our study of deconfinement and
chiral symmetry restoration at finite temperature in a strong
magnetic background. This study is motivated by several reasons.
Firstly, it is extremely interesting to understand how an external
field can modify the main characteristics of confinement and
spontaneous chiral symmetry breaking. Lattice studies on QCD in
magnetic (as well as chromo-magnetic) backgrounds can be found
in~\cite{D'Elia:2010nq,Cea:2002wx,Buividovich:2009my}. Studies of
QCD in magnetic fields, and of QCD-like theories as well, can be
found in Refs.~\cite{Klevansky:1989vi,
Fukushima:2010fe,Mizher:2010zb,Gatto:2010qs,Campanelli:2009sc,Chernodub:2010qx,
Frolov:2010wn,Bergman:2008sg}. Besides, strong magnetic fields
might be produced in non-central heavy ion
collisions~\cite{Kharzeev:2007jp,Skokov:2009qp}. More concretely,
at the center-of-mass energy reachable at LHC, $\sqrt{s_{NN}}
\approx 4.5$ TeV, the magnetic field can be as large
as~\footnote{In this article, we measure $eB$ in units of the
vacuum squared pion mass $m_\pi^2$; then, $eB=m_\pi^2$ corresponds
to $B\approx 2\times 10^{14}$ Tesla. } $eB \approx 15 m_\pi^2$
according to~\cite{Skokov:2009qp}. It has been argued that in
these conditions, the sphaleron transitions of finite temperature
QCD, give rise to Chiral Magnetic Effect (CME)
\cite{Kharzeev:2007jp,Buividovich:2009wi}.

The novelty of this study is the use of the EPNJL model in our
calculations, in the one-loop approximation. In comparison with
the original PNJL model of~\cite{Fukushima:2003fw}, the EPNJL
model has two additional parameters. However, they are fixed from
the QCD thermodynamics at zero magnetic field, as we will discuss
in more detail later. Therefore, the results at a finite value of
the magnetic field strength have to be considered as predictions
of the model. Our main result is that the entanglement of the NJL
coupling constant and the Polyakov loop, might affect crucially
the phase diagram in the temperature/magnetic field strength
plane. Previous model
studies~\cite{Fukushima:2010fe,Mizher:2010zb,Gatto:2010qs} have
revealed that both the deconfinement temperature, $T_L$, and the
chiral symmetry restoration temperature, $T_\chi$, are enhanced by
a magnetic field, in agreement with the Lattice data of
Ref.~\cite{D'Elia:2010nq}. The model results are in slight
disagreement with the Lattice, in the sense that the former
predict a considerable split of $T_L$ and $T_\chi$ as the strength
of the magnetic field is increased. Within the EPNJL model, we can
anticipate one of the results, that is, the split among $T_L$ and
$T_\chi$ might be considerably reduced even at large values of the
magnetic field strength. In particular, using the values of the
parameters of~\cite{Sakai:2010rp}, which arise from a best-fit of
Lattice data at zero and imaginary chemical potential and which
are appropriate for our study, we find a split of the order of
$2\%$ at the largest value of $eB$ considered, namely $eB = 30
m_\pi^2$.

In~\cite{Gatto:2010qs} a similar computation within a model
without entanglement, but with an 8-quark interaction added, has
been performed. Following the nomenclature of~\cite{Sakai:2010rp},
we call the latter model, PNJL$_8$ model. We find the comparison
among the EPNJL and the PNJL$_8$ models very instructive. As a
matter of fact, the parameters in the two models are chosen to
reproduce the QCD thermodynamics at zero and imaginary chemical
potential~\cite{Sakai:2010rp,Sakai:2009dv}. Therefore, both of
them are capable to describe QCD in the same regime. It is
interesting that, because of the different interactions content,
the two models predict a slight different behavior of hot quark
matter in strong magnetic field. In the next future, the
comparison with refined Lattice data can enlighten on which of the
two models is a more faithful description of QCD.

The paper is organized as follows: in Section II we summarize the
formalism: we derive the quark propagator and the equation for the
chiral condensates in magnetic field, within the EPNJL model, in
the one-loop approximation; then, we compute the thermodynamic
potential. in Section III, we collect our results for the chiral
condensate and the expectation value of the Polyakov loop, for
several values of the magnetic field strength. In Section IV, we
briefly investigate on the effect of the 8-quark interaction in
the EPNJL model in magnetic field. In Section V, we draw the phase
diagram of the EPNJL model in magnetic field, and make a
comparison with our previous result~\cite{Gatto:2010qs}. Finally,
in Section VI we draw our conclusions, and briefly comment on
possible extensions and prosecutions of our study.

We use natural units throughout this paper, $\hbar = c = k_B = 1$,
and work in Euclidean space-time $R^4 = \beta V$ , where $V$ is
the volume and $\beta = 1/T$ with $T$ corresponding to the
temperature of the system. Moreover, we take a non-zero current
quark mass. In this case, both deconfinement and chiral symmetry
breaking are crossovers; however, we sometimes will make use of
the term ``phase transition'' to describe them, for stylistic
reasons. It should be clear from the context that our ``phase
transitions'' are meant to be crossovers unless stated
differently.

\section{The model with entanglement vertex}
We consider a model in which quark interaction is described by the
following lagrangian density:
\begin{equation}
{\cal L} =\bar\psi\left(i\gamma^\mu D_\mu - m\right)\psi + {\cal
L}_I~;\label{eq:1ooo}
\end{equation}
here $\psi$ is the quark Dirac spinor in the fundamental
representation of the flavor $SU(2)$ and the color group;
$\bm\tau$ correspond to the Pauli matrices in flavor space. A sum
over color and flavor is understood. The covariant derivative
embeds the QED coupling of the quarks with the external magnetic
field, as well as the QCD coupling with the background gluon field
which is related to the Polyakov loop, see below. Furthermore, we
have defined
\begin{equation}
{\cal L}_I = G\left[\left(\bar\psi\psi\right)^2 +
\left(i\bar\psi\gamma_5\bm\tau\psi\right)^2\right]~;\label{eq:1}
\end{equation}
This interaction term is invariant under $SU(2)_V\otimes
SU(2)_A\otimes U(1)_V$. In the chiral limit, this is the symmetry
group of the action as well, if no magnetic field is applied.
However, this group is broken explicitly to $U(1)_V^3\otimes
U(1)_A^3\otimes U(1)_V$ if the magnetic field is coupled to the
quarks, because of the different electric charge of $u$ and $d$
quarks. Here, the superscript $3$ in the $V$ and $A$ groups
denotes the transformations generated by $\tau_3$,
$\tau_3\gamma_5$ respectively. Therefore, the chiral group in
presence of a magnetic field is $U(1)_V^3\otimes U(1)_A^3$. This
group is then explicitly broken by the quark mass term to
$U(1)_V^3$.

We are interested to the interplay among chiral symmetry
restoration and deconfinement in a strong magnetic field. To
compute a temperature for the deconfinement crossover, we use the
expectation value of the Polyakov loop, that we denote by $L$. In
order to compute $L$ we introduce a static, homogeneous and
Euclidean background temporal gluon field, $A_0 = iA_4 = i
\lambda_a A_4^a$, coupled minimally to the quarks via the QCD
covariant derivative~\cite{Fukushima:2003fw}. Then
\begin{equation}
L = \frac{1}{3}\text{Tr}_c\exp\left(i\beta\lambda_a A_4^a\right)~,
\end{equation}
where $\beta = 1/T$. In the Polyakov gauge, which is convenient
for this study, $A_0 = i\lambda_3 \phi + i \lambda_8 \phi^8$;
moreover, we work at zero quark chemical potential, therefore we
can take $L = L^\dagger$ from the beginning, which implies $A_4^8
= 0$. This choice is also motivated by the study
of~\cite{Mizher:2010zb}, where it is shown that the paramagnetic
contribution of the quarks to the thermodynamic potential induces
the breaking of the $Z_3$ symmetry, favoring the configurations
with a real-valued Polyakov loop (see Section III.C
of~\cite{Mizher:2010zb} for an excellent discussion of this
point). We have then~\cite{Fukushima:2003fw,Ratti:2005jh}
\begin{equation}
L = \frac{1+2\cos(\beta\phi)}{3}~.
\end{equation}

As already discussed in the Introduction, it has been shown that
it is possible to derive the effective 4-quark
interaction~\eqref{eq:1} starting from the QCD
lagrangian~\cite{Kondo:2010ts}. In~\cite{Kondo:2010ts} it has been
shown that the NJL vertex has a non-local structure, that is, it
is momentum-dependent. An analogous conclusion is achieved
in~\cite{Frasca:2008zp}. More important for our study, the NJL
vertex acquires a non-trivial dependence on the phase of the
Polyakov loop. Therefore, in the model we consider here, it is
important to keep into account this dependence. The exact
dependence of $G$ on $L$ has not yet been computed; it is possible
that it will be determined in the next future by means of the
functional renormalization group approach~\cite{Kondo:2010ts}. In
our study, we follow a more phenomenological approach to the
problem, using the ansatz introduced in~\cite{Sakai:2010rp}, that
is
\begin{equation}
G = g\left[1 - \alpha_1 L\bar{L} -\alpha_2(L^3 +
\bar{L}^3)\right]~,\label{eq:Run}
\end{equation}
and we take $L=\bar{L}$ from the beginning. The functional form in
the above equation is constrained by $C$ and extended $Z_3$
symmetry. We refer to~\cite{Sakai:2010rp} for a more detailed
discussion. The numerical values of $\alpha_1$ and $\alpha_2$ have
been fixed in~\cite{Sakai:2010rp} by a best fit of the available
Lattice data at zero and imaginary chemical potential of
Ref.~\cite{D'Elia:2009qz}, which have been confirmed recently
in~\cite{Bonati:2010gi}. In particular, the fitted data are the
critical temperature at zero chemical potential, and the
dependence of the Roberge-Weiss endpoint on the bare quark mass.

The values $\alpha_1 = \alpha_2 \equiv \alpha = 0.2 \pm 0.05$ have
been obtained in~\cite{Sakai:2010rp} using a hard cutoff
regularization scheme. We will focus mainly on the case
$\alpha=0.2$ as in~\cite{Sakai:2010rp}. We have verified that in
our regularization scheme, our results are in quantitative
agreement with those of~\cite{Sakai:2010rp}, when we take the
parameter $T_0 = 190$ MeV in the Polyakov loop effective potential
in agreement with~\cite{Sakai:2010rp}, see below. Then, we will
study how the results change when we vary $\alpha$.

\subsection{Quark propagator and chiral condensate in magnetic field}
We work in the mean field approximation throughout this paper,
neglecting pseudoscalar condensates; moreover, we make the
assumption that condensation takes place only in the flavor
channels $\tau_0$ and $\tau_3$. The mean field interaction term
Eq.~\eqref{eq:1} can be cast in the form
\begin{equation}
{\cal L} = -2G\Sigma \left(\bar u u + \bar d d\right)-G
\Sigma^2~,\label{eq:4}
\end{equation}
where $\Sigma = -\langle\bar{u}u + \bar{d}d\rangle$.

In a magnetic field, the chiral condensates of $u$ and $d$ quarks
have to be different, because the electric charges of these quarks
are different. Even if the one-loop quark self-energies in
Eq.~\eqref{eq:4} depend on the sum of the two condensates, being
therefore flavor independent, it is straightforward to show that
the two condensates turn out to be different, by taking the trace
of the propagator of the two quarks. The interaction~\eqref{eq:4}
is diagonal in flavor space, therefore we can focus on the
propagator of a single flavor $f$.

To write the quark propagator we use the Ritus
method~\cite{Ritus:1972ky}, which allows to expand the propagator
on the complete and orthonormal set made of the eigenfunctions of
a charged fermion in a homogeneous and static magnetic field. This
is a well known procedure, discussed many times in the literature,
see for
example~\cite{Elizalde:2000vz,Ferrer:2005vd,Fukushima:2009ft,Fukushima:2007fc};
therefore it is enough to quote the final result,
\begin{equation}
S_f(x,y) = \sum_{k=0}^\infty\int\frac{dp_0 dp_2 dp_3}{(2\pi)^4}
E_P(x)\Lambda_k \frac{1}{P\cdot\gamma - M}\bar{E}_P(y)~,
\label{eq:QP}
\end{equation}
where $E_P(x)$ corresponds to the eigenfunction of a charged
fermion in magnetic field, and $\bar{E}_P(x) \equiv
\gamma_0(E_P(x))^\dagger \gamma_0$. In the above equation,
\begin{equation}
P = (p_0 + i A_4,0,{\cal Q}\sqrt{2k|Q_f eB|},p_z)~,\label{eq:MB}
\end{equation}
where $k =0,1,2,\dots$ labels the $k^{\text{th}}$ Landau level,
and ${\cal Q} \equiv\text{sign}(Q_f)$, with $Q_f$ denoting the
charge of the flavor $f$; $\Lambda_k$ is a projector in Dirac
space which keeps into account the degeneracy of the Landau
levels; it is given by
\begin{equation}
\Lambda_k = \delta_{k0}\left[{\cal P_+}\delta_{{\cal Q},+1} +
{\cal P_-}\delta_{{\cal Q},-1}\right] + (1-\delta_{k0})I~,
\end{equation}
where ${\cal P}_{\pm}$ are spin projectors and $I$ is the identity
matrix in Dirac spinor indices. The propagator in
Eq.~\eqref{eq:QP} has a non-trivial color structure, due to the
coupling to the background gauge field, see Eq.~\eqref{eq:MB}.

The trace of the $f$-quark propagator is minus the chiral
condensate $\langle\bar f f\rangle$, with $f=u,d$. Taking the
trace in coordinate and internal space, it is easy to show that
the following equation holds:
\begin{eqnarray}
\langle\bar f f\rangle &=& - N_c\frac{|Q_f
eB|}{2\pi}\sum_{k=0}^\infty \beta_k \int\frac{dp_z}{2\pi}
\frac{M_f}{\omega_f} {\cal C}(L,\bar L, T|p_z,
k)~.\nonumber\\
&&\label{eq:CC}
\end{eqnarray}
Here,
\begin{eqnarray}
{\cal C}(L,\bar L, T|p_z, k) &=& U_\Lambda - 2{\cal N}(L,\bar L,
T|p_z, k)~,\label{eq:CCc}
\end{eqnarray}
and ${\cal N}$ denotes the statistically confining Fermi
distribution function,
\begin{eqnarray}
{\cal C}(L,\bar L, T|p_z, k) &=& \frac{1 + 2L e^{\beta\omega_f} +
Le^{2\beta\omega_f} }
  {1+3L e^{\beta\omega_f} + 3L e^{2\beta\omega_f} +
  e^{3\beta\omega_f}}~,\nonumber\\
  &&
\end{eqnarray}
where
\begin{equation}
\omega_f^2 = p_z^2 + 2|Q_f e B|k + M_f^2~,
\end{equation}
with
\begin{equation}
M_u =M_d =m_0 +2G \Sigma~. \label{eq:MAsse}
\end{equation}
The first and the second addenda in the r.h.s. of
Eq.~\eqref{eq:CC} correspond to the vacuum and the thermal
fluctuations contribution to the chiral condensate, respectively.
The coefficient $\beta_k = 2 -\delta_{k0}$ keeps into account the
degeneracy of the Landau levels. The vacuum contribution is
ultraviolet divergent. In order to regularize it, we adopt a
smooth regulator $U_\Lambda$, which is more suitable, from the
numerical point of view, in our model calculation with respect to
the hard-cutoff which is used in analogous calculations without
magnetic field. In this article we chose
\begin{equation}
U_\Lambda = \frac{\Lambda^{2N}}{\Lambda^{2N} + (p_z^2 + 2|Q_f e
B|k)^N}~.
\end{equation}
To be specific, we consider here the case $N=5$, for numerical
convenience. A larger value of $N$ increases the cutoff artifacts,
as already discussed in detail
in~\cite{Campanelli:2009sc,Fukushima:2010fe,Gatto:2010qs}, but
leaves the qualitative picture unchanged; on the other hand, a
smaller value of $N$, namely $N \leq 3$, makes not possible the
fit of the pion decay constant and of the chiral condensate in the
vacuum. The (more usual) 3-momentum cutoff regularization scheme
is recovered in the limit $N\rightarrow\infty$; we notice that,
even if the choice $N=5$ may seem arbitrary to some extent, it is
not more arbitrary than the choice of the hard cutoff, that is, of
a regularization scheme.

By virtue of Eq.~\eqref{eq:CC} it is easy to argue that the
condensates of $u$ and $d$ quarks are different in magnetic field.
As a matter of fact, the different value of the charges makes the
r.h.s. of the above equation flavor-dependent, even if $\Sigma$ is
flavor-singlet. Once we compute the mean field value of $\Sigma$
by minimization of the thermodynamic potential (see the next
Section), we will use Eq.~\eqref{eq:CC} to evaluate the chiral
condensates for $u$ and $d$ quarks in magnetic field.

\subsection{Thermodynamic potential}
In the one-loop approximation, the thermodynamic potential
$\Omega$ is given by
\begin{equation}
\Omega ={\cal U}(L,\bar L,T) + G\Sigma^2 -\frac{1}{\beta
V}\text{Tr}\log\left(\beta S^{-1}\right)~,
\end{equation}
where the trace is over color, flavor, Dirac and space-time
indices and the propagator for each flavor is given by
Eq.~\eqref{eq:QP}. It is straightforward to derive the final
result, namely
\begin{eqnarray}
\Omega &=& {\cal U}(L,\bar L,T) +U_M  \nonumber \\
&&-\sum_{f=u,d}\frac{|Q_f
eB|}{2\pi}\sum_{k}\beta_k\int_{-\infty}^{+\infty}\frac{dp_z
}{2\pi} {\cal G}(L,\bar L,T|p_z,k)~.\nonumber\\
&&~\label{eq:OB}
\end{eqnarray}
In the above equation, the first line corresponds to the classical
contribution; we have defined
\begin{equation}
U_M = G_S \Sigma^2~;
\end{equation}
the second line corresponds to the sum of vacuum and thermal one
loop contributions, respectively, and arise after integration over
the fermion fluctuations in the functional integral. We have
defined
\begin{equation}
{\cal G}(L,\bar L, T|p_z,k) = N_c U_\Lambda(\bm p)\omega_f +
\frac{2}{\beta}\log{\cal F}~,
\end{equation}
with
\begin{equation}
{\cal F}(L,\bar L, T|p_z,k) = 1 +3L e^{-\beta\omega_f}+3\bar{L}
e^{-2\beta\omega_f}+e^{-3\beta\omega_f}~.
\end{equation}

The potential term $\mathcal{U}[L,\bar L,T]$ in Eq.~\eqref{eq:OB}
is built by hand in order to reproduce the pure gluonic lattice
data~\cite{Ratti:2005jh,Roessner:2006xn}. We adopt the following
logarithmic form~\cite{Roessner:2006xn},
\begin{equation}
 \begin{split}
 & \mathcal{U}[L,\bar L,T] = T^4\biggl\{-\frac{a(T)}{2}
  \bar L L \\
 &\qquad + b(T)\ln\bigl[ 1-6\bar LL + 4(\bar L^3 + L^3)
  -3(\bar LL)^2 \bigr] \biggr\} \;,
 \end{split}
\label{eq:Poly}
\end{equation}
with three model parameters (one of four is constrained by the
Stefan-Boltzmann limit),
\begin{equation}
 \begin{split}
 a(T) &= a_0 + a_1 \left(\frac{T_0}{T}\right)
 + a_2 \left(\frac{T_0}{T}\right)^2 , \\
 b(T) &= b_3\left(\frac{T_0}{T}\right)^3 \;.
 \end{split}
\label{eq:lp}
\end{equation}
The standard choice of the parameters reads $a_0 = 3.51$, $a_1 =
-2.47$, $a_2 = 15.2$ and $b_3 = -1.75$. The parameter $T_0$ in
Eq.~\eqref{eq:Poly} sets the deconfinement scale in the pure gauge
theory. In absence of dynamical fermions one has $T_0 = 270$
\text{MeV}. However, dynamical fermions induce a dependence of
this parameter on the number of active
flavors~\cite{Herbst:2010rf}. For the case of two light flavors to
which we are interested here, we take $T_0 = 212$ MeV as
in~\cite{Gatto:2010qs}.

As a final remark, it is easy to show that summing
Eq.~\eqref{eq:CC} for $u$ and $d$ quarks, one reproduces the
equation satisfied by $\Sigma$, the latter being obtained from the
stationarity condition $\partial\Omega/\partial\Sigma = 0$.

\section{Deconfinement and chiral symmetry restoration}
In this Section we summarize our numerical results. In this study
we fix the values of $g$, $\Lambda$ and $m_0$ to reproduce the
values of $f_\pi$ and $m_\pi$ in the vacuum, as well as the
numerical value of the light quarks chiral condensate. We have
then $\Lambda = 626.76$ MeV, $g =2.02/\Lambda^2$. The pion mass in
the vacuum, $m_\pi = 139$ MeV, is used to fix the numerical value
of the bare quark mass via the GMOR relation $f_\pi^2 m_\pi^2 = -2
m_0\langle\bar u u\rangle$; this gives $m_0 = 5$ MeV when we take
$\langle\bar u u\rangle = (-253~\text{MeV})^3$. Finally, we take
$T_0 = 212$ MeV in the Polyakov loop effective potential as
in~\cite{Gatto:2010qs}, unless stated differently. The values of
$\Sigma$ and $L$ are computed by the minimization of the
thermodynamic potential.

\begin{figure}
\begin{center}
\includegraphics[width=7cm]{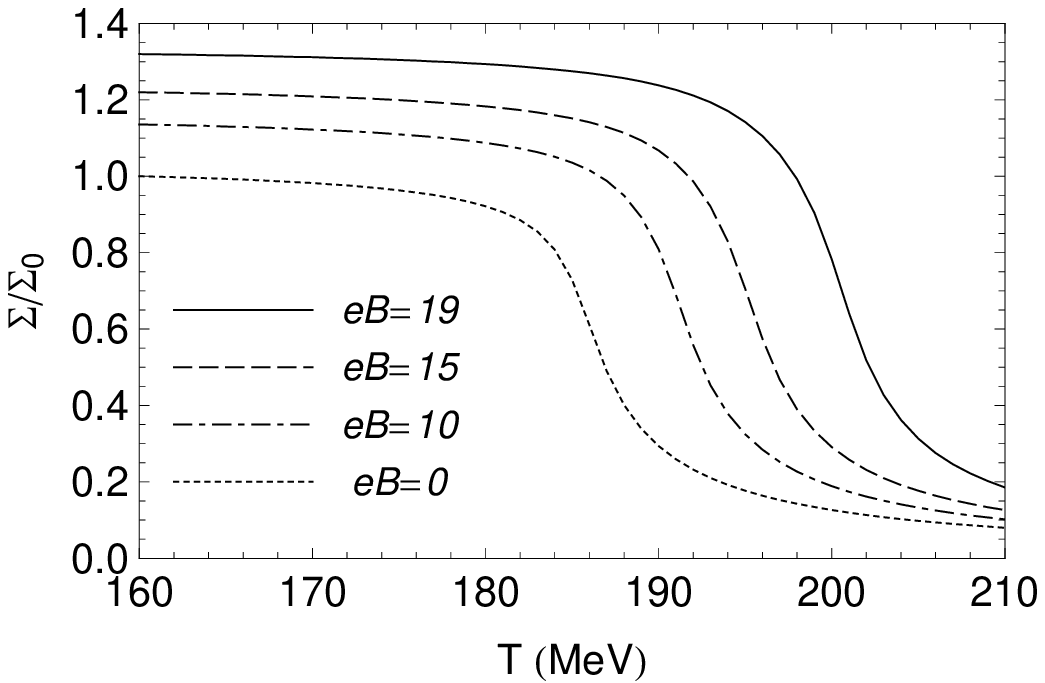}\\
\includegraphics[width=7cm]{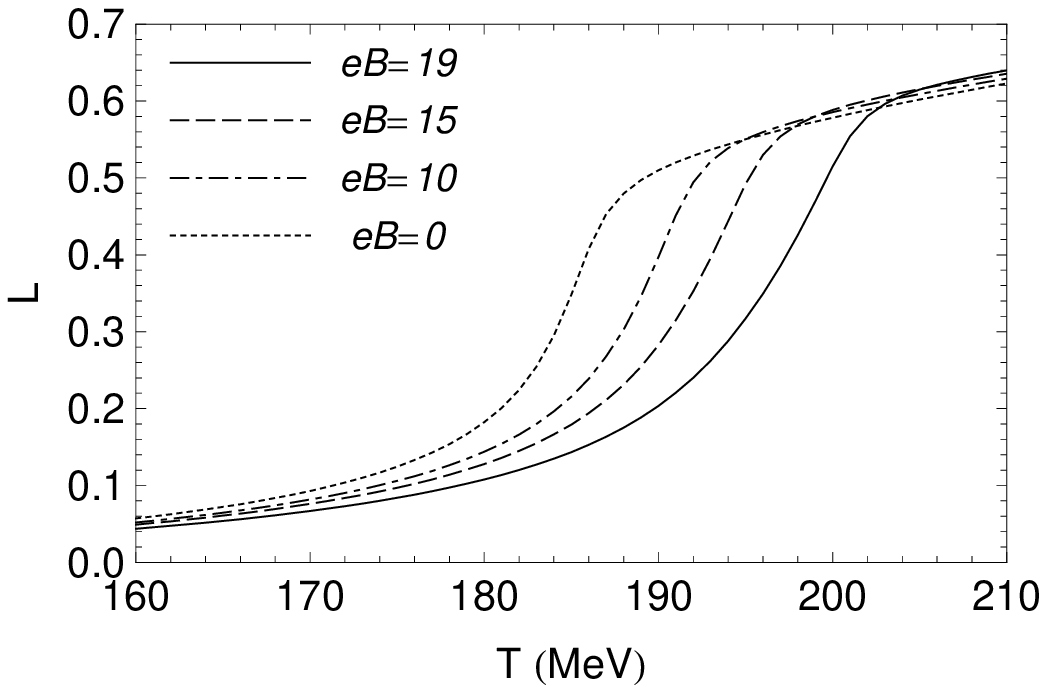}
\end{center}
\caption{\label{Fig:cond} {\em Upper panel:} Chiral condensate
$\Sigma$, measured in units of the chiral condensate at zero
temperature and zero field, $\Sigma_0$, as a function of
temperature, for several values of the magnetic field strength.
{\em Lower panel:} Expectation value of the Polyakov loop, $L$, as
a function of temperature, for several values of the magnetic
field strength. The data are obtained for $\alpha=0.2$. In the
figures, the magnetic fields are measured in units of $m_\pi^2$.}
\end{figure}

\subsection{The case $\alpha_1 = \alpha_2 = 0.2$}
In Fig.~\ref{Fig:cond} we plot our data for the chiral condensate
$\Sigma$, measured in units of the condensate at $T=eB=0$, that is
$\Sigma_0 = 2\times(-253~\text{MeV})^3$, and the expectation value
of the Polyakov loop as a function of temperature, computed for
several values of the magnetic field strength. They are computed
from the minimization of the thermodynamic potential in
Eq.~\eqref{eq:OB}. In the numerical computation, we have chosen
$\alpha_1 = \alpha_2 \equiv \alpha = 0.2$ as
in~\cite{Sakai:2010rp}.

The results shown in Fig.~\ref{Fig:cond} are interesting for
several reasons. Firstly, if we identify the deconfinement
crossover with the temperature $T_L$ at which $dL/dT$ is maximum,
and the chiral crossover with the temperature $T_\chi$ at which
$|d\Sigma/dT|$ is maximum, we observe that the two temperatures
are very close also in a strong magnetic field. For concreteness,
at $eB=0$ we find $T_P = T_\chi = 185.5$ MeV. Besides, at $eB=19
m_\pi^2$ we find $T_P =199$ MeV and $T_\chi = 201$ MeV. Hence at
the strongest value of $eB$ considered here, namely $eB=19
m_\pi^2$, the entanglement vertex makes the split of the two
crossovers of $\approx 1.5\%$. From the model point of view, it is
easy to understand why deconfinement and chiral symmetry
restoration are entangled also in strong magnetic field. As a
matter of fact, using the data shown in Fig.~\ref{Fig:cond}, it is
possible to show that the NJL coupling constant in the
pseudo-critical region in this model decreases of the $15\%$ as a
consequence of the deconfinement crossover. Therefore, the
strength of the interaction responsible for the spontaneous chiral
symmetry breaking is strongly affected by the deconfinement, with
the obvious consequence that the numerical value of the chiral
condensate drops down and the chiral crossover takes place. We
have verified that the picture remains qualitatively and
quantitatively unchanged if we perform a calculation at $eB=30
m_\pi^2$. In this case, we find $T_L = 224$ MeV and $T_\chi = 225$
MeV.

Before going ahead, it worths to notice that we have also
considered the case $T_0 = 190$ MeV, which corresponds to the
value considered in~\cite{Sakai:2010rp}. In this case, for $eB=0$
we find $T_\chi = T_L = 175$ MeV, in excellent agreement
with~\cite{Sakai:2010rp}. This results is comforting, because it
shows that even using a different regularization scheme, the
UV-regulator does not affect the physical predictions at $eB=0$.
Moreover, at $eB/m_\pi^2 = 30$, we find $T_\chi \approx T_L = 215$
MeV.

This result can be compared with our previous
calculation~\cite{Fukushima:2010fe} in which we did not include
the Polyakov loop dependence of the NJL coupling constant.
In~\cite{Fukushima:2010fe} we worked in the chiral limit and we
observed that the Polyakov loop crossover in the PNJL model is
almost insensitive to the magnetic field; on the other hand, the
chiral phase transition temperature was found to be very sensitive
to the strength of the applied magnetic field, in agreement with
the well known magnetic catalysis
scenario~\cite{Klevansky:1989vi}. This model prediction has been
confirmed within the Polyakov extended quark-meson model
in~\cite{Mizher:2010zb}, when the contribution from the vacuum
fermion fluctuations to the energy density is kept into
account~\footnote{If the vacuum corrections are neglected, the
deconfinement and chiral crossovers are found to be coincident
even in very strong magnetic fields~\cite{Mizher:2010zb}, but the
critical temperature decreases as a function of $eB$; this
scenario is very interesting theoretically, but it seems it is
excluded from the recent Lattice
simulations~\cite{D'Elia:2010nq}.}; we then obtained a similar
result in~\cite{Gatto:2010qs}, in which we turned from the chiral
to the physical limit at which $m_\pi = 139$ MeV, and introduced
the 8-quark term as well (PNJL$_8$ model, according to the
nomenclature of~\cite{Sakai:2010rp}). The comparison with the
results of the PNJL$_8$ model of~\cite{Gatto:2010qs} is
interesting because the model considered there, was tuned in order
to reproduce the Lattice data at zero and imaginary chemical
potential~\cite{Sakai:2009dv}, like the model we use in this
study. Therefore, they share the property to describe the QCD
thermodynamics at zero and imaginary chemical potential; it is
therefore instructive to compare their predictions at finite $eB$.

For concreteness, in~\cite{Gatto:2010qs} we found $T_P =185$ MeV
and $T_\chi = 208$ MeV at $eB=19 m_\pi^2$, corresponding to a
split of $\approx 12\%$. On the other hand, in the present
calculation we measure a split of $\approx 1.5\%$ at the largest
value of $eB$ considered. Therefore, the results of the two models
are in slight quantitative disagreement; this disagreement is then
reflected in a slightly different phase diagram. We will draw the
phase diagram of the two models in a next Section; however, since
now it is easy to understand what the main difference consists in:
the PNJL$_8$ model predicts some window in the $eB-T$ plane in
which chiral symmetry is still broken by a chiral condensate, but
deconfinement already took place. In the case of the EPNJL model,
this window is shrunk to a very small one, because of the
entanglement of the two crossovers at finite $eB$. On the other
hand, it is worth to stress that the two models share an important
qualitative feature: both chiral restoration and deconfinement
temperatures are enhanced by a strong magnetic field, in
qualitative agreement with the existing Lattice
data~\cite{D'Elia:2010nq}.

A further interesting feature of our data is that in the low
temperature region, the Polyakov loop is suppressed as we increase
the strength of the magnetic field; on the other hand, in the
deconfinement phase, $L$ is enhanced by the magnetic field. This
result was also found in our previous
studies~\cite{Fukushima:2010fe,Gatto:2010qs} with a fixed coupling
constant; more remarkably, this result is in agreement with the
Lattice data~\cite{D'Elia:2010nq}. At the moment we do not have
physical arguments to explain this result, but we agree with the
authors of Ref.~\cite{D'Elia:2010nq} that this aspect should be
considered in more detail from the theoretical point of view.

\begin{figure}
\begin{center}
\includegraphics[width=7cm]{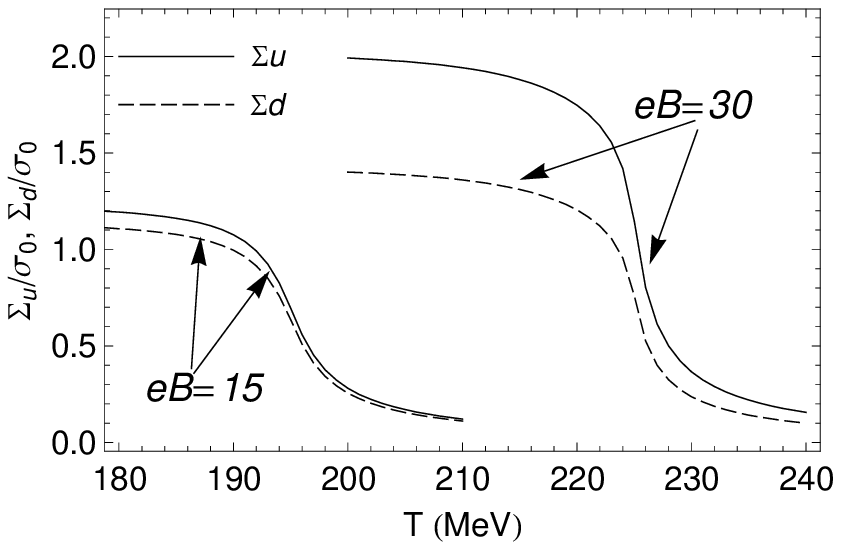}\\
\includegraphics[width=7cm]{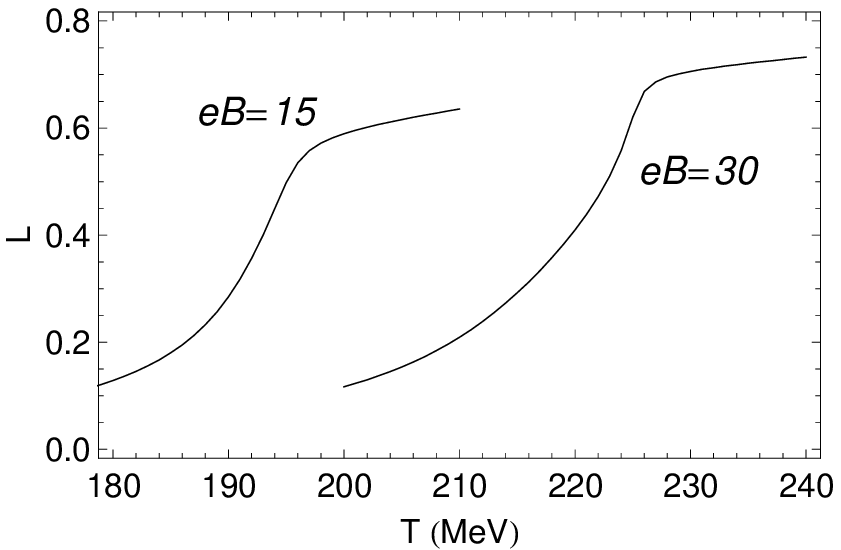}
\end{center}
\caption{\label{Fig:UDCC} {\em Upper panel:} Chiral condensates of
$u$ and $d$ quarks as functions of temperatures in the
pseudo-critical region, at $eB = 15 m_\pi^2$ and $eB = 30
m_\pi^2$. Condensates are measured in units of their value at zero
magnetic field and zero temperature, namely $\sigma_0 =
(-253~\text{MeV})^3$. {\em Lower panel:} Polyakov loop expectation
value as a function of temperature, at $eB = 15 m_\pi^2$ and $eB =
30 m_\pi^2$. Data correspond to $\alpha=0.2$.}
\end{figure}

For completeness, we plot in Fig~\ref{Fig:UDCC} the chiral
condensates of $u$ and $d$ quarks as a function of temperature, at
$eB = 15 m_\pi^2$ and $eB = 30 m_\pi^2$. The condensates are
measured in units of their value at zero magnetic field and zero
temperature, namely $\sigma_0 \equiv \langle\bar uu\rangle =
\langle\bar dd\rangle = (-253~\text{MeV})^3$. They are computed by
a two-step procedure: firstly we find the values of $\Sigma$ and
$L$ that minimize the thermodynamic potential; then, we make use
of Eq.~\eqref{eq:CC} to compute the expectation values of $\bar
uu$ and $\bar dd$ in magnetic field. If we measure the strength of
the crossover by the value of the peak of $|d\Sigma/dT|$, it is
obvious from the Figure that the chiral crossover becomes stronger
and stronger as the strength of the magnetic field is increased,
in agreement with~\cite{D'Elia:2010nq}.

\subsection{Varying $\alpha$}

\begin{figure}
\begin{center}
\includegraphics[width=7cm]{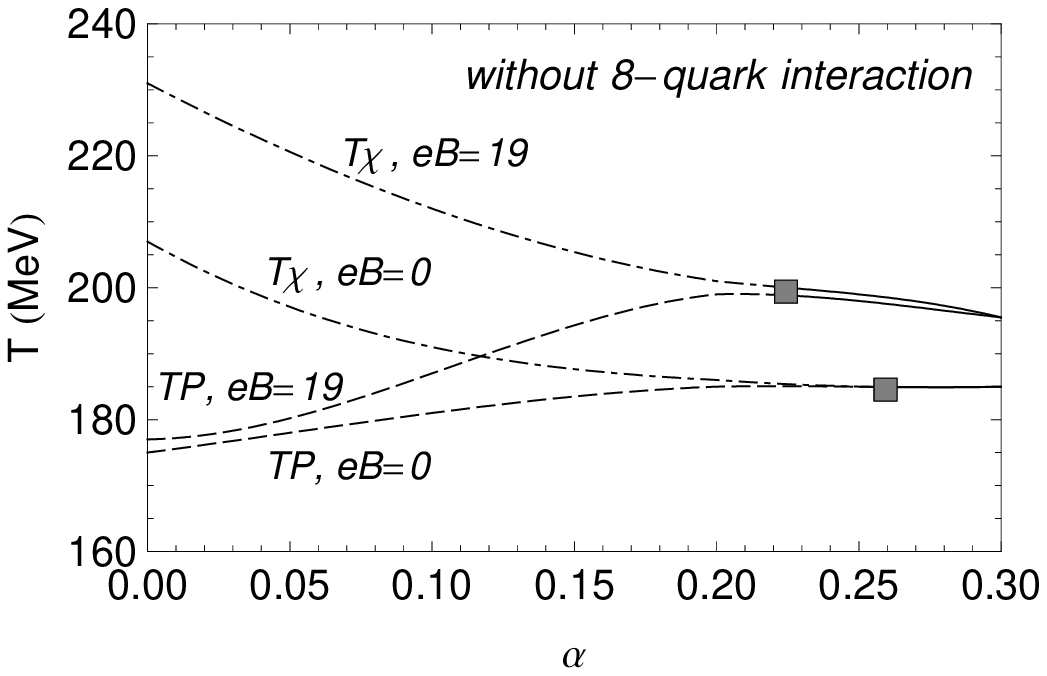}\\
\includegraphics[width=7cm]{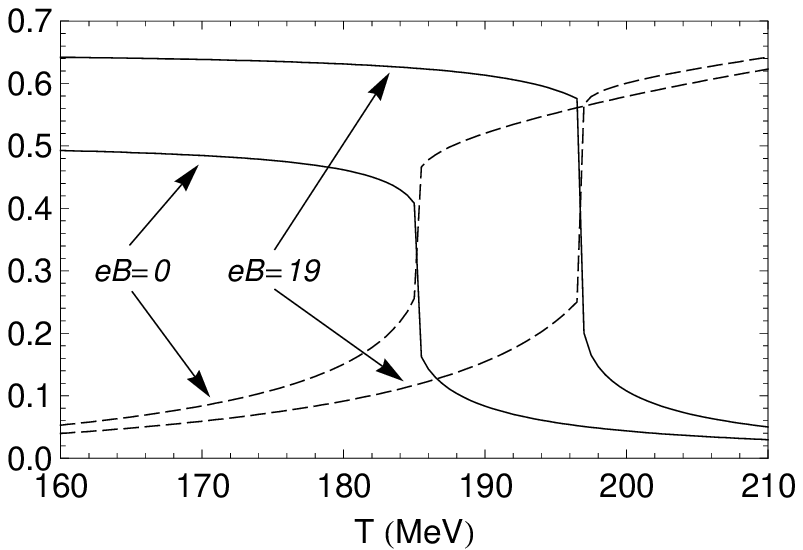}
\end{center}
\caption{\label{Fig:alpha} {\em Upper panel:} pseudo-critical
temperatures as functions of the parameter $\alpha$ in the NJL
coupling constant. Dashed lines correspond to the Polyakov loop
crossover, for $eB=0$ and $eB=19$. Dot-dashed lines correspond to
the chiral crossover. The grey squares denote the value of
$\alpha$ at which the crossover becomes a first-order phase
transition. {\em Lower panel:} Polyakov loop expectation value
(dashed lines) and half of $\Sigma(eB)/\Sigma(eB=0)$ (solid lines)
as functions of temperature, for $\alpha = 0.275$. In both panels
the magnetic fields are measured in units of $m_\pi^2$.}
\end{figure}

Since our study is based on a simple model, and we can not compute
the exact values of the parameter $\alpha$ in Eq.~\eqref{eq:Run}
starting from first principles, we find very instructive to
measure the robustness of our results as we vary the numerical
value of this parameter. We emphasize that in~\cite{Sakai:2010rp}
an estimate for the numerical value of $\alpha$ is achieved via
the best fit of the model data with the Lattice data at zero and
imaginary chemical potential. However, in that reference a
different regularization scheme is used. Therefore, we can expect
that a slightly different value of $\alpha$ is needed in our case.
However, it is useful to observe that in the limit of zero field,
our regulator is very similar, quantitatively speaking, to the
hard-cutoff of~\cite{Sakai:2010rp}, because of the strong
power-law decay of $U_\Lambda$ as $|\bm p|\approx\Lambda$.
Therefore, the possible numerical deviation in the parameters is
expected to be very tiny, if any. We have a further corroboration
we are on target with parameters: if we chose $T_0 = 190$ MeV as
in~\cite{Sakai:2010rp}, we find $T_\chi = T_L = 175$ MeV, in
quantitative agreement with~\cite{Sakai:2010rp}; this convinces
that our ultraviolet regulator does not lead to quantitative
discrepancy with~\cite{Sakai:2010rp} at zero magnetic field.

We have investigated the dependence of the pseudo-critical
temperatures as a function of $\alpha$, at $eB=0$ and $eB=19
m_\pi^2$. The results of our computations are collected in the
upper panel of Fig.~\ref{Fig:alpha}. At $\alpha=0$, which
corresponds to the original PNJL model, we measure a split of the
two critical temperatures of $\approx 18\%$ even at $eB=0$. This
is well known result in the PNJL
literature~\cite{Roessner:2006xn,Ratti:2005jh}, and to overcome
this problem, several solutions have been suggested, like the use
of the 8-quark
interaction~\cite{Sakai:2010rp,Bhattacharyya:2010wp}. However, the
discrepancy of the critical temperatures is enhanced as the value
of $eB$ is increased. At $eB=19 m_\pi^2$ we find a split of
$\approx 30\%$. As we increase $\alpha$, the two temperatures get
closer rapidly, both at zero and at non-zero value of the magnetic
field strength. At $\alpha=0.2$, which is the result that we have
discussed previously and is the value quoted
in~\cite{Sakai:2010rp}, the two temperatures are almost
coincident.

An interesting point that we have found is that increasing further
the value of $\alpha$, the crossovers become stronger and
stronger; there exists a critical value of $\alpha$ at which the
crossover is replaced by a sudden jump in the chiral condensate
and in the Polyakov loop expectation value. This value of $\alpha$
is denoted by a green square in Fig.~\ref{Fig:alpha}, and it is
$eB-$dependent. For concreteness, in the lower panel of
Fig.~\ref{Fig:alpha} we plot our data for the half of the chiral
condensate and the Polyakov loop at $\alpha=0.275$, which should
be compared with the data in Fig.~\ref{Fig:cond} where
$\alpha=0.2$.

\section{The effect of the 8-quark interaction}
We have briefly investigated on the role of other interactions on
the entanglement of deconfinement and chiral symmetry restoration
crossovers. We report here the results related to the effect of
the 8-quark term, which has been extensively studied
in~\cite{Osipov:2005tq} in relation to the vacuum stability of the
NJL model, to the hadron properties and to the critical
temperature at zero baryon density; besides, it has been studied
in the context of the P-NJL model
in~\cite{Sakai:2010rp,Bhattacharyya:2010wp,Gatto:2010qs}. To this
end, we add to the interaction lagrangian in Eq.~\eqref{eq:1}, the
following term
\begin{equation}
G_8\left[\left(\bar\psi\psi\right)^2 +
\left(i\bar\psi\gamma_5\bm\tau\psi\right)^2\right]^2~.\label{eq:1bis}
\end{equation}
In principle, we expect that the constant $G_8$ acquires a
dependence on the Polyakov loop as well; however, for simplicity
we neglect this dependence here, since we are only interested to
understand the qualitative effect of the
interaction~\eqref{eq:1bis}. We remark that the model studied in
this Section is different from the P-NJL$_8$ model defined
in~\cite{Sakai:2010rp}; in the latter, the dependence of $G$ on
$L$ is neglected.

At the one-loop order we have
\begin{eqnarray}
{\cal L} &=& -2\bar u u \left[G \Sigma + 2G_8\Sigma^3\right] -
2\bar dd \left[G \Sigma  + 2G_8\Sigma^3\right] \nonumber
\\
&&-3G_8\Sigma^4-G \Sigma^2~,\label{eq:4bis}
\end{eqnarray}
instead of Eq.~\eqref{eq:4}. The thermodynamic potential is still
given by Eq.~\eqref{eq:OB}, with
\begin{equation}
U_M = 3G_8\Sigma^4 +G_S \Sigma^2~,
\end{equation}
and
\begin{equation}
M_u = M_d = m_0 +2(G \Sigma + 2G_8 \Sigma^3)~, \label{eq:MAsseBIS}
\end{equation}
instead of Eq.~\eqref{eq:MAsse}. The parameters are fixed as
in~\cite{Gatto:2010qs}, that is $\Lambda= 589$ MeV, $g=5\times
10^{-6}$ MeV$^{-2}$, $G_8 = 6\times 10^{-22}$ MeV$^{-8}$ and $m_0
= 5.6$ MeV.

\begin{figure}
\begin{center}
\includegraphics[width=7cm]{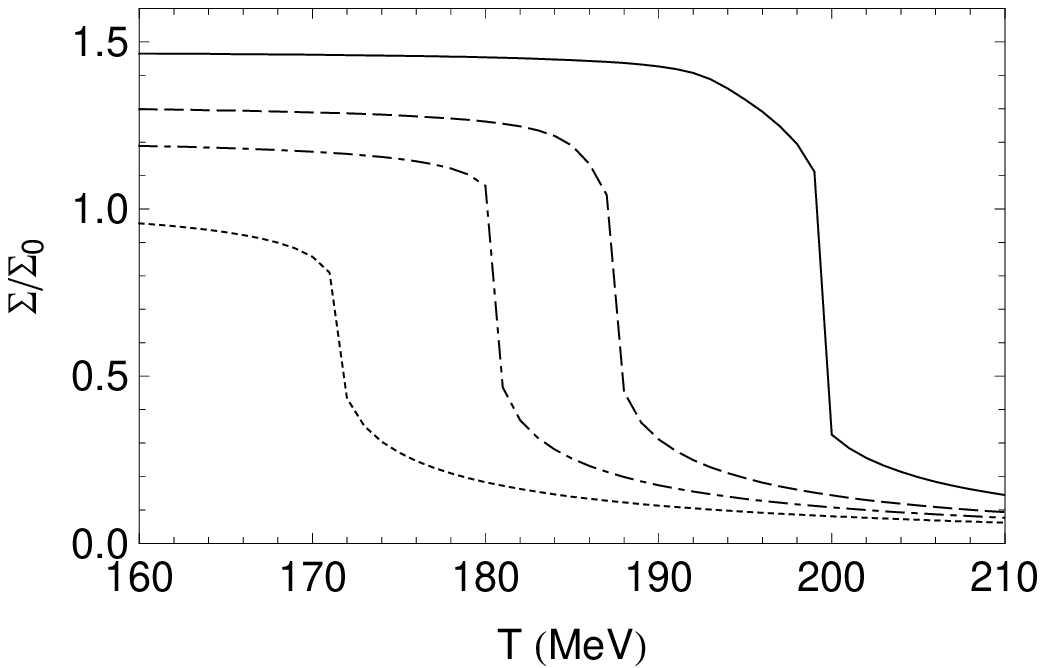}\\
\includegraphics[width=7cm]{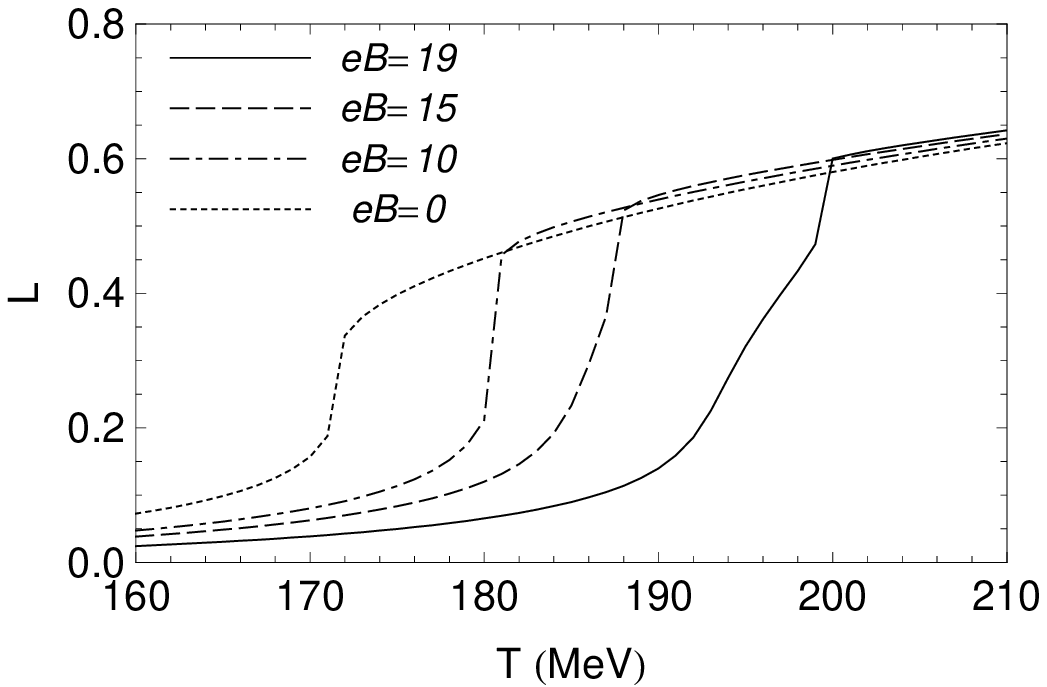}
\end{center}
\caption{\label{Fig:cond8} {\em Upper panel:} Chiral condensate
$\Sigma$, measured in units of the chiral condensate at zero
temperature and zero field, $\Sigma_0$, as a function of
temperature, for several values of the magnetic field strength.
{\em Lower panel:} Expectation value of the Polyakov loop $P$ as a
function of temperature, for several values of the magnetic field
strength. The results are obtained within the model with 8-quark
interaction and $\alpha=0.1$. In the figures, the magnetic fields
are measured in units of $m_\pi^2$. Lines with the same dashing
correspond to the same value of the magnetic field strength.}
\end{figure}

In Fig.~\ref{Fig:cond8} we plot our data for the chiral condensate
and the expectation value of the Polyakov loop, as a function of
temperature, for several values of the magnetic field strength and
for $\alpha=0.1$. We find that even at this smaller value of
$\alpha$, the two crossovers are entangled at large magnetic
fields. On the other hand, we find that taking a running NJL
coupling makes the crossovers sharper, and eventually we measure a
discontinuity in the order parameters for a large magnetic field
strength.

\begin{figure}
\begin{center}
\includegraphics[width=7cm]{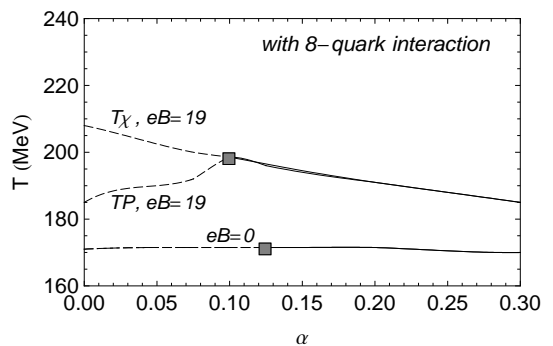}
\end{center}
\caption{\label{Fig:alpha8} Pseudo-critical temperatures as
functions of the parameter $\alpha$ in the NJL coupling constant.
Dashed lines correspond to the Polyakov loop crossover, for $eB=0$
and $eB=19$. Dot-dashed lines correspond to the chiral crossover.
The grey squares denote the value of $\alpha$ at which the
crossover becomes a first-order phase transition. }
\end{figure}

In Fig.~\ref{Fig:alpha8} we plot the critical temperatures as a
function of $\alpha$, for $eB=0$ and $eB=19m_\pi^2$. At $eB=0$ the
two crossovers are entangled already at $\alpha=0$; therefore, the
effect of the entanglement vertex is to make the crossover
sharper. Eventually, at the critical value $\alpha = 0.125$ the
crossover is replaced by a sudden jump of the expectation values,
analogously to the results that we find in the model without
8-quark term. As we increase $eB$, the deconfinement and chiral
crossovers are split at $\alpha=0$, in agreement with our previous
findings~\cite{Gatto:2010qs}. However, the split in this case is
more modest than the split that we measure in the model with
$G_8=0$. In that case, at $eB=19 m_\pi^2$ we find a split of
$\approx 30\%$, to be compared with the one that we read from
Fig.~\ref{Fig:alpha8} for the model with 8-quark interaction,
namely $\approx 12\%$. Therefore, our conclusion is that the
8-quark interaction helps to keep the two crossovers close in a
strong magnetic field, even if it is not enough and the
entanglement vertex has to be included, as it is clear from
Fig.~\ref{Fig:alpha8}.

\section{The phase diagram} 
\begin{figure}
\begin{center}
\includegraphics[width=7cm]{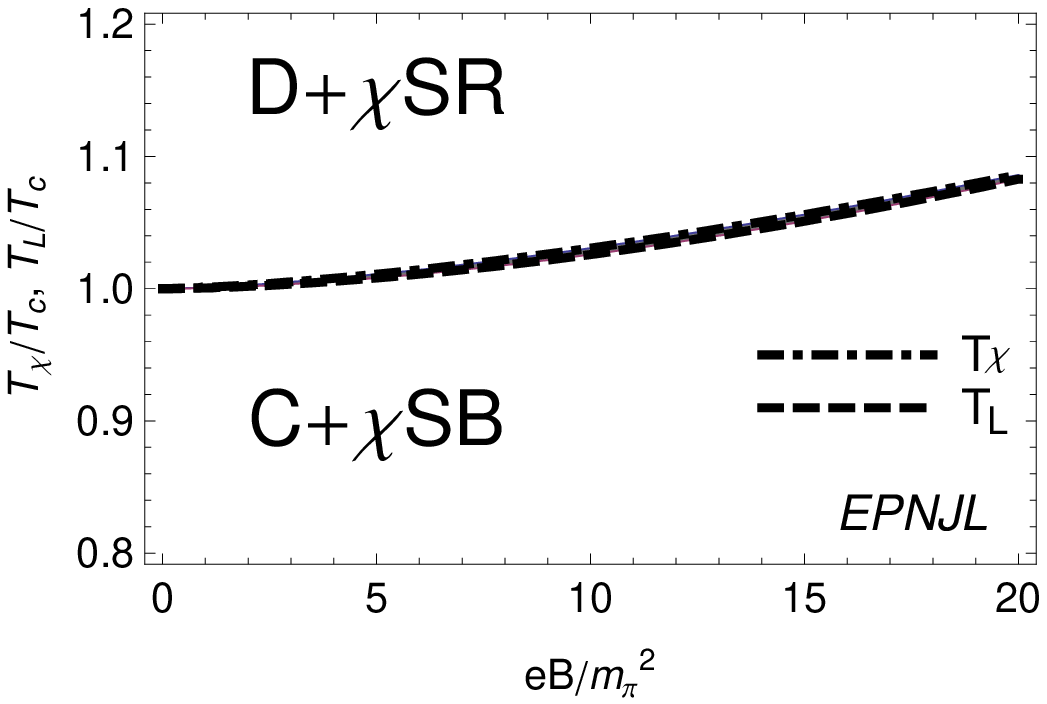}\\
\includegraphics[width=7cm]{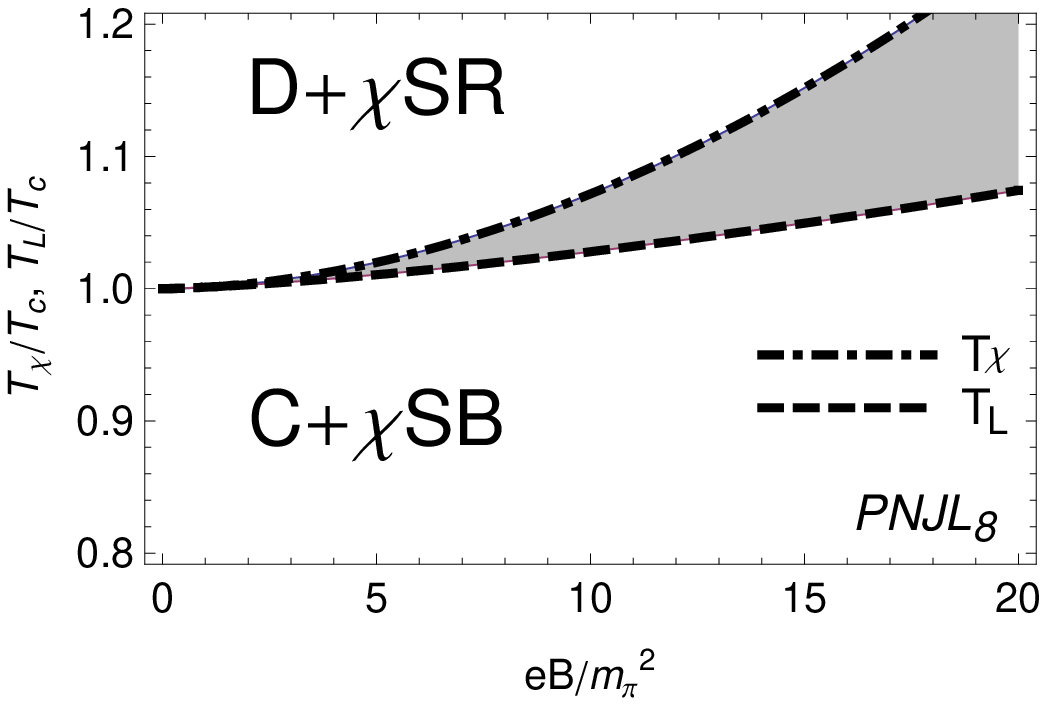}
\end{center}
\caption{\label{Fig:PD2} {\em Upper panel:} Phase diagram in the
$eB-T$ plane for the EPNJL model. Temperatures on the vertical
axis are measured in units of the pseudo-critical temperature for
deconfinement at $eB=0$, namely $T_c = 185.5$ MeV. {\em Lower
panel:} Phase diagram in the $eB-T$ plane for the PNJL$_8$ model.
Temperatures on the vertical axis are measured in units of the
pseudo-critical temperature for deconfinement at $eB=0$, namely
$T_c = 175$ MeV. In both the phase diagrams, $T_\chi$, $T_L$
correspond to the chiral and deconfinement pseudo-critical
temperatures, respectively. The grey shaded region denotes the
portion of phase diagram in which hot quark matter is deconfined
and chiral symmetry is still broken spontaneously.}
\end{figure}

In Fig.~\ref{Fig:PD2} we collect our
data on the pseudo-critical temperatures for deconfinement and
chiral symmetry restoration, in the form of a phase diagram in the
$eB-T$ plane. In the upper panel we show the results obtained
within the EPNJL model; in the lower panel, we plot the results of
the PNJL$_8$ model, that are obtained using the fitting functions
computed in~\cite{Gatto:2010qs}. In the figure, the magnetic field
is measured in units of $m_\pi^2$; temperature is measured in
units of the deconfinement pseudo-critical temperature at zero
magnetic field, namely $T_{B=0} = 185.5$ MeV for the EPNJL model,
and $T_{B=0} = 175$ MeV for the PNJL$_8$ model. For any value of
$eB$, we identify the pseudo-critical temperature with the peak of
the effective susceptibility.

It should be kept in mind, however, that the definition of a
pseudo-critical temperature in this case is not unique, because of
the crossover nature of the phenomena that we describe. Other
satisfactory definitions include the temperature at which the
order parameter reaches one half of its asymptotic value (which
corresponds to the $T\rightarrow 0$ limit for the chiral
condensate, and to the $T\rightarrow +\infty$ for the Polyakov
loop), and the position of the peak in the true susceptibilities.
The expectation is that the critical temperatures computed in
these different ways differ from each other only of few percent.
This can be confirmed concretely using the data in
Fig.~\ref{Fig:UDCC} at $eB=30 m_\pi^2$. Using the peak of the
effective susceptibility we find $T_\chi = 225$ MeV and $T_L =
224$ MeV; on the other hand, using the half-value criterion, we
find $T_\chi = 227$ MeV and $T_L = 222$ MeV, in very good
agreement with the previous estimate. Therefore, the qualitative
picture that we derive within our simple calculational scheme,
namely the entanglement of the two crossovers in a strong magnetic
field, should not be affected by using different definitions of
the critical temperatures.

Firstly we focus on the phase diagram of the EPNJL model, which is
one of the novelties of our study. In the upper panel of
Fig.~\ref{Fig:PD2}, the dashed and dot-dashed lines correspond to
the deconfinement and chiral symmetry restoration pseudo-critical
temperatures, respectively. We have fit our data using the
following functional form:
\begin{equation}
\frac{T_{\chi,L}(eB)}{T_c} = 1 + A
\left(\frac{|eB|}{T_c^2}\right)^\alpha~,\label{eq:fitTc}
\end{equation}
where the subscripts $\chi, L$ correspond to the (approximate)
chiral restoration and deconfinement temperatures, respectively.
The functional form is taken in analogy to the one used
in~\cite{D'Elia:2010nq}. Numerical values of the best-fit
coefficients are collected in Table~\ref{Tab:fit}. As a
consequence of the entanglement, the two crossovers stay closed
also in very strong magnetic field, as we have already discussed
in the previous Sections. The grey region in the Figure denotes a
phase in which quark matter is (statistically) deconfined, but
chiral symmetry is still broken. According
to~\cite{Cleymans:1986cq,Kouno:1988bi}, we can call this phase as
Constituent Quark Phase (CQP).

\begin{table}[t!]
\caption{\label{Tab:fit}Coefficients of the fit function defined
in Eq.~\eqref{eq:fitTc} for the two models.}
\begin{ruledtabular}
\begin{tabular}{cccc}
 &$A$& $\alpha$ & $T_c$ (MeV) \\
\hline

$T_\chi$, EPNJL &$2.34\times 10^{-3}$ &1.49 & 185.5 \\
\hline

$T_L$, EPNJL  &$1.43\times 10^{-3}$ &1.68 & 185.5 \\
\hline

$T_\chi$, PNJL$_8$~\cite{Gatto:2010qs} &$2.4\times 10^{-3}$ &1.85 & 175 \\
\hline

$T_L$, PNJL$_8$~\cite{Gatto:2010qs}  &$2.1\times 10^{-3}$ &1.41 & 175 \\

\end{tabular}
\end{ruledtabular}
\end{table}

On the lower panel of Fig.~\ref{Fig:PD2} we have drawn the phase
diagram for the PNJL$_8$ model; it is obtained using
Eq.~\eqref{eq:fitTc} with the coefficients computed in
Ref.~\cite{Gatto:2010qs}, and collected in Table~\ref{Tab:fit}.
The critical temperature at zero field in this case is a little
smaller than the same temperature for the EPNJL model, because the
parameters needed to fit the vacuum properties are different.
However, a comparison  among the two phase diagrams is still
instructive. The most astonishing feature of the phase diagram of
the PNJL$_8$ model is the entity of the split among the
deconfinement and the chiral restoration crossover. The difference
with the result of the EPNJL model is that in the former, the
entanglement with the Polyakov loop is neglected in the NJL
coupling constant. As we have already mentioned in the previous
Section, the maximum amount of split that we find within the EPNJL
model, at the largest value of magnetic field considered here, is
of the order of $2\%$; this number has to be compared with the
split at $eB=20 m_\pi^2$ in the PNJL$_8$ model, namely $\approx
12\%$. The larger split causes a considerable portion of the phase
diagram to be occupied by the CQP.

Keeping into account the crudeness of the two models, it is fair
to say that they share one important feature: both deconfinement
and chiral symmetry restoration temperature are increased by a
strong magnetic field. They disagree quantitatively, in the sense
that the amount of split in the EPNJL model is more modest in
comparison with that obtained in the PNJL$_8$ model. Before
closing, we notice that a similar phase diagram calculation has
been performed within the Quark-Meson (QM in the following) model
in~\cite{Mizher:2010zb}. In this reference, the magnetic field has
the same effect that we measure in the PNJL$_8$ model, that is, to
split of deconfinement and chiral symmetry restoration crossovers.
Besides, both $T_\chi$ and $T_L$ are enhanced by a magnetic field,
in excellent agreement with the behavior that we find within the
several NJL-like models that we have used in this study.

On the Lattice side, the most recent data about this kind of study
are those of Ref.~\cite{D'Elia:2010nq}. In these data, the split
among the two crossovers is absent. Then the Lattice data seem to
point towards the phase diagram of the EPNJL model. On the other
hand, the results of~\cite{D'Elia:2010nq} might not be definitive,
in the sense that both the lattice size might be enlarged, and the
pion mass could be lowered to its physical value in the vacuum.
Therefore, it will be interesting to compare our results with more
refined data.

\section{Conclusions}
In this article, we have studied chiral symmetry restoration and
deconfinement in a strong magnetic background, using an effective
model of QCD. In particular, we have reported our results about
the effect of the entangled vertex on the phase diagram. Our main
result is that the entanglement reduces considerably the split
among the deconfinement and the chiral symmetry crossovers studied
in~\cite{Fukushima:2010fe,Mizher:2010zb,Gatto:2010qs}, as
expected.

We have also studied the effect of the 8-quark term on the split.
Our results suggest that the 8-quark interaction helps the two
crossovers to be close. Furthermore, we have shown that the
crossovers become sharper and then they are replaced by a sudden
jump of the expectation values, as the value of $\alpha$ in the
entanglement vertex is larger than a critical value.

We have then compared our results with those of other model
calculations, namely the PNJL model~\cite{Fukushima:2010fe}, the
Quark-Meson model~\cite{Mizher:2010zb} and the PNJL$_8$
model~\cite{Gatto:2010qs}. The most striking similarity among the
several models is that they all support the scenario in which
chiral symmetry restoration and deconfinement temperatures are
enhanced by a strong magnetic field. The models differ
quantitatively for the amount of split measured (very few percent
for the EPNJL model, and of the order of $10\%$ for the other
models for $eB \approx 20 m_\pi^2$).

Furthermore, we have compared our results with those obtained on
the Lattice~\cite{D'Elia:2010nq}. In~\cite{D'Elia:2010nq}, the
largest value of magnetic field considered is $eB\approx 0.75$
GeV$^2$, which corresponds to $eB/m_\pi^2 \approx 38$. The Lattice
data seem to point towards the phase diagram of the EPNJL model.
On the other hand, the results of~\cite{D'Elia:2010nq} might not
be definitive: the lattice size might be enlarged, the lattice
spacing could be taken smaller (in~\cite{D'Elia:2010nq} the
lattice spacing is $a=0.3$ fm), and the pion mass could be lowered
to its physical value in the vacuum. As a consequence, it will be
interesting to compare our results with more refined data in the
future.

This comparison can be interesting also for another reason.
Indeed, the EPNJL model and the PNJL$_8$ can describe the same QCD
thermodynamics at zero and imaginary quark chemical
potential~\cite{Sakai:2010rp}, but they differ qualitatively for
the interaction content. As we have shown, they have some
quantitative discrepancy for what concerns the response to a
strong magnetic field. Therefore, more refined Lattice data in
magnetic field might help to discern which of the two models is a
more faithful description of QCD.

From our point of view, it is fair to admit that our study might
have a weak point, namely, we miss a microscopic computation of
the parameters $\alpha_1$, $\alpha_2$ in Eq.~\eqref{eq:Run}. For
concreteness, we have used the best-fit values quoted
in~\cite{Sakai:2010rp}, showing in addition that changing the
value of $T_0$ in the Polyakov loop effective potential as
in~\cite{Sakai:2010rp}, we obtain $T_\chi = T_L = 175$ MeV, in
excellent agreement  with that reference. This is comforting,
since it shows that our different UV-regulator does not affect the
qualitative result at zero field, namely the coincidence of
$T_\chi$ and $T_L$. Besides, we are aware that Eq.~\eqref{eq:Run}
is just a particular choice of the functional dependence of the
NJL coupling constant on the Polyakov loop expectation value.
Different functional forms respecting the $C$ and the extended
$Z_3$ symmetry are certainly possible, and without a rigorous
derivation of Eq.~\eqref{eq:Run} using functional renormalization
group techniques as suggested in~\cite{Kondo:2010ts}, it merits to
study our problem using different choices for $G(L)$ in the next
future. For these reasons, we prefer to adopt a conservative point
of view: it is interesting that a model which is adjusted in order
to reproduce Lattice data at zero and imaginary chemical
potential, predicts that the two QCD transitions are entangled in
a strong magnetic background; however, this conclusion might not
be definitive, since there exist other model calculations which
share a common basis with ours, and which show a more pronounced
split of the QCD transitions in a strong magnetic background. More
refined Lattice data will certainly help to discern which of the
two scenarios is the most favorable.

As a natural continuation of this work, it is worth to perform the
computation of the chiral magnetization~\cite{Buividovich:2009my}
at finite temperature. Besides, the technical machinery used here
can be easily applied to a microscopic study of the spectral
properties of mesons in strong magnetic field. In this direction,
it is interesting to compute the masses of the charged
$\rho$-mesons at low temperature, in order to investigate their
condensation at large magnetic field as suggested
in~\cite{Chernodub:2010qx}. Furthermore, it would be interesting
to make a complete (analytical or semi-analytical) study of the
chiral limit, to estimate the effect of the magnetic field on the
universality class of two-flavor QCD. We will report on these
topics in the next future.

\acknowledgments We acknowledge E.~Cartman, M.~Cristoforetti,
A.~Flachi, K.~Fukushima, F.~Greco, E.~M.~Ilgenfritz, T.~Kahara,
K.~I.~Kondo, V.~Mathieu and H.~Warringa, for valuable discussions
and encouraging comments about the topics discussed in this
article. Besides, we acknowledge E.~Fraga, A.~Mizher and
A.~Ohnishi for a careful reading of the manuscript. Moreover, we
acknowledge many discussions and heavy correspondence with
M.~Chernodub, M. D'Elia and M.~Frasca. We also thank T.~Kunihiro
for comments about the effects of the 8-quark term on the phase
transitions. The work of M.~R.\ is supported by JSPS under the
contract number P09028. The numerical calculations were carried
out on Altix3700 BX2 at YITP in Kyoto University.



\end{document}